\documentclass[12pt]{article}
\usepackage[a4paper,top=2.5cm,bottom=2.5cm,left=2.5cm,right=2.5cm,marginparwidth=1cm]{geometry}
\usepackage{times}
\usepackage{setspace}
\usepackage[document]{ragged2e}
\usepackage{amsmath}
\usepackage{graphicx}
\usepackage{wrapfig}
\usepackage[utf8]{inputenc}
\usepackage[english]{babel}
\usepackage[sorting=none]{biblatex}
\usepackage[font=small]{caption}
\usepackage{subfig}
\usepackage{xcolor}
\usepackage{hyperref}

\begin{document}
\begin{titlepage}
    \begin{center}
        \vspace*{1cm}
 
        \Huge
        \textbf{Simulation of Muon Tomography Projections to Image the Pyramids of Giza}
 
        \vspace{0.5cm}
        \large
        Presentation of a simulation of geometric tracks of muon trajectories through a pyramid to a two-plane detector used in one sided muon tomography
    \end{center}

        \noindent
        {Author: Mira Liu \\ 
        Advisor: Patrick La Riviere, PhD}\\
        
        \vspace{0.5cm}
        
        \noindent
        Submitted as a rotation report presented for the fulfillment of University of Chicago Medical Physics MPHY 41800 \textit{Research in Advanced Tomographic Imaging} with Dr. La Riviere.

        \vspace{0.5cm}
        \noindent
        Work presented here is not peer-reviewed and is not submitted for independent publication. It is part of a project proposed by an interdisciplinary group of researchers from the Oriental Institute, the University of Virginia, Fermilab, the Geophysics Department of Cairo University, and the University of Chicago Medical Physics collaborating to produce high-resolution tomographic images. Work presented here contributed to published paper\footnote{Bross, Alan David, Dukes, E. C., Ehrlich, Ralf, Fernandez, Eric, Dukes, Sophie, Gobashy, Mohamed, Jamieson, Ishbel, La Riviere, Patrick J., Liu, Mira, Marouard, Gregory, Moeller, Nadine, Pla-Dalmau, Anna, Rubinov, Paul, Shohoud, Omar, Vargas, Phillip, and Welch, Tabitha. 2022. "Tomographic Muon Imaging of the Great Pyramid of Giza". United States. https://doi.org/10.31526/jais.2022.280.} available here: \href{https://www.osti.gov/biblio/1844785}{https://www.osti.gov/biblio/1844785}
    \begin{center}
        \vfill
        \noindent
        \large
        Graduate Program in Medical Physics\\
        University of Chicago\\
        Chicago, IL\\
        April 4, 2020
    \end{center}
    
\end{titlepage}
\section{Abstract}

\indent\indent\textbf{Purpose: } A geometric simulation of a possible two-plane detector was developed to test the abilities of the detector to generate high-resolution images of the Great Pyramid using muon tomography.

\textbf{Methods and Materials:} Trajectory range, angular resolution, and acceptance of the detector were calculated with a simulation. Trajectories and the corresponding sinogram space covered were simulated first with one detector in one location, and then two moving detectors on adjacent sides of the pyramid. The resolution at the center slice of the pyramid was calculated using the angular resolution of the detector.

\textbf{Results:} The simulation returned trajectory range encompassing the pyramid and peak angular resolution of $.0004 sr$. Sinogram space covered by one position was inadequate, however two moving detectors on adjacent sides of the pyramid cover a significant portion. Resolution at the center of the pyramid is roughly $3m$.

\textbf{Conclusions:} The simulation provides a way to calculate the detector positions needed to cover an adequate amount of sinogram space for high-resolution cosmic-ray tomographic reconstruction of the Great Pyramids.

\textbf{Key Words:} high-resolution muon tomography, one-sided tomography, sinogram simulation, detector simulation

\section{Introduction}

Knowledge of the Great Pyramid of Giza is currently limited to knowledge from optical instruments. The question of how these structures differ, and reasons why, call for detailed, high-resolution imaging of those internal core structures by other means. In 2017 a new void was detected by the Scan Pyramids team through cosmic-ray muon tomography, albeit with relatively poor resolution \cite{Morishima}. A project has since been proposed by an interdisciplinary group of researchers from the Oriental Institute, the University of Virginia, Fermilab, the Geophysics Department of Cairo University, and the University of Chicago Medical Physics collaborating to produce high-resolution tomographic images. These would reconstruct the interior structure of the Great Pyramid of Giza with one-sided cosmic-ray muon tomography\cite{Bross}. The Medical Physics aspect involves exploring the ability of the proposal to meet the required tomographic sampling necessary for a high-resolution 3D reconstruction of the pyramid. This rotation project focused on building a simulation to study the geometric dimensions of the proposed detector and the Great Pyramid, angular resolution and acceptance of the detector, resolution of detection, and the range of sinogram space able to be covered. 

\section{Methods and Materials}
The dimensions of the proposed detectors and the dimensions of the external shape of the Great Pyramid were taken from a previous GEANT4 monte carlo simulation of cosmic-ray muon tomography\cite{Ehrlich}. A numerical simulation was written to show that the widest trajectory angles subtended the entire Great Pyramid if the far detector side was located 25m from the edge of the pyramid. Given the pixel dimensions of the detector, the angular resolution and acceptance of the detectors were calculated using the definitions provided by Lesparre et. al.\cite{Lesparre} From this, a simulation was written to translate all potential trajectories of the detector to sinogram space covered to test the ability of high-resolution tomographic reconstruction. This simulation can be used to determine placement and movement of the detectors at the base of the pyramid. Lastly, the simulation was extended to know the resolution at the furthest depth of the pyramid given the angular resolution of the pixels. As this is a purely geometric simulation, the cosmic ray spectrum, probability of interaction, energy of incident muon, scattering, density, and attenuation were not included. These sections are explored in detail in the subsections below.\footnote{All code was written in Python3 and source code developed is available at:\\ https://github.com/miramliu/MuonTomography.git} 

\subsection{Geometric Dimensions}

The proposed detector is built within a shipping container, of length $9.6m$, height $2.4m$, and depth $2m$\cite{Ehrlich}. A front panel and a back panel of scintillator modules allow hits to be connected to define a muon's trajectory through the detector. On each panel $2 \times 2 cm^2$ pixels are arranged such that there are $120$ pixels vertically, and $480$ pixels horizontally. This creates a grid of $240\times480 = 115200$ ij-pixels (i representing the horizontal location, j representing the vertical location), each of which have a trajectory through the matching $115200$ ij-pixels $D = 2m$ away on the opposite panel. This generates $1.3\times 10^{10}$ paths between the two panels. Trajectories of muons are calculated by reconstructing the vector connecting the pixel in the front array and back array knowing the distance between the two pixels $m=\Delta i$ and $n=\Delta j$ with $\Vec{v} = [\Delta i,\Delta j]$ as shown in Figure\ref{fig:DetectorDiagram}. The trajectories with the greatest angle provide the limit of the angular range. The pyramid with dimensions of depth and width $230.33m$ and height $138.7m$ must fit within this angular range to be imaged in its entirety.

\subsection{Angular Resolution and Acceptance}

Given the dimensions and number of the detector pixels, the angular resolution and acceptance can be calculated. Using the definition from Lesparre et. al.\cite{Lesparre} the angular resolution for each discrete direction of sight of the detector is equal to the solid angle covering all trajectories able to hit a given pair of pixels of that direction of sight. This angular resolution was simulated with and without tilt correction. This angular resolution and the area of the detector with this angular resolution were used to calculate the acceptance.

\begin{figure}[ht]
    \centering
    \begin{minipage}[t]{.39\textwidth}
        \includegraphics[width=1\textwidth]{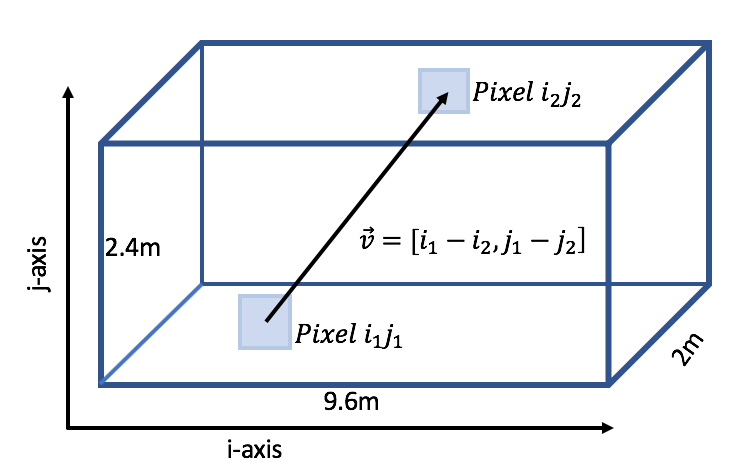}
        \captionsetup{width=1.1\linewidth, format=hang}
        \caption{\label{fig:DetectorDiagram}\noindent A diagram of the shipping container detector proposed using the $ij$ coordinate system. A trajectory and its vector is shown.}
    \end{minipage}
    \hfill
    \begin{minipage}[t]{.29\textwidth}
        \includegraphics[width=1\textwidth]{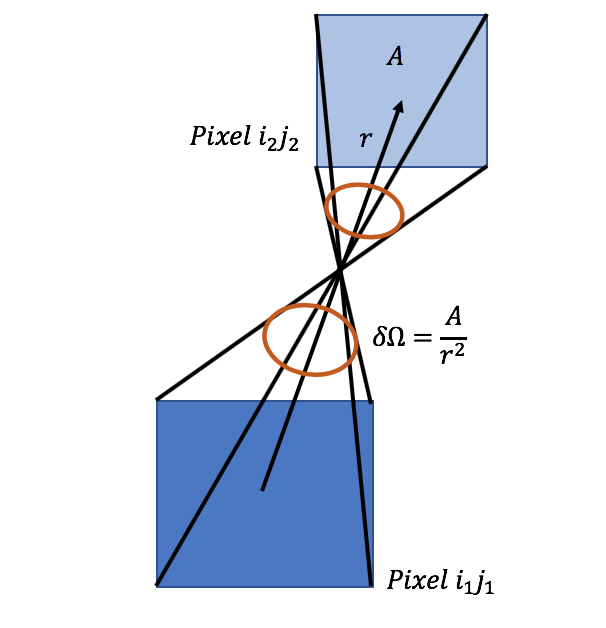}
        \captionsetup{width=1.1\linewidth, format=hang}
        \caption{\label{fig:SolidAngleDiagram}A diagram of the un-tilted solid angle and calculation used in the code.}
    \end{minipage}
    \hfill
    \begin{minipage}[t]{.29\textwidth}
        \includegraphics[width=1\textwidth]{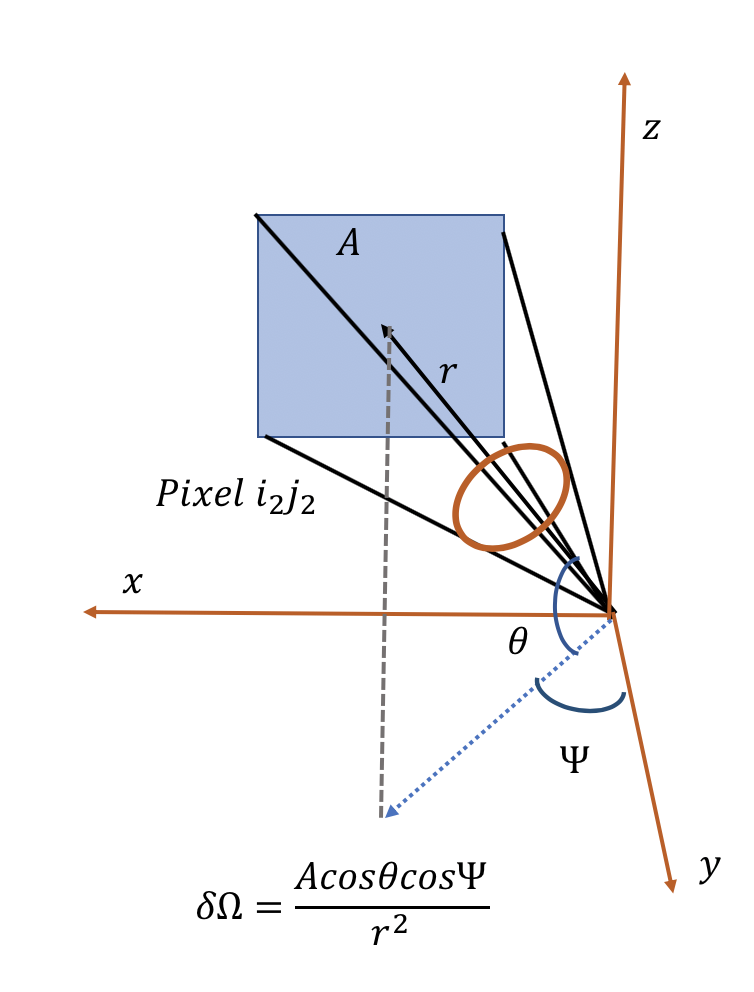}
        \captionsetup{width=1.1\linewidth, format=hang}
        \caption{\label{fig:TiltedSolidAngleDiagram}A diagram of the tilted solid angle and the calculation used in the code.}
    \end{minipage}
    \begin{minipage}[t]{.44\textwidth}
        \includegraphics[width=1\textwidth]{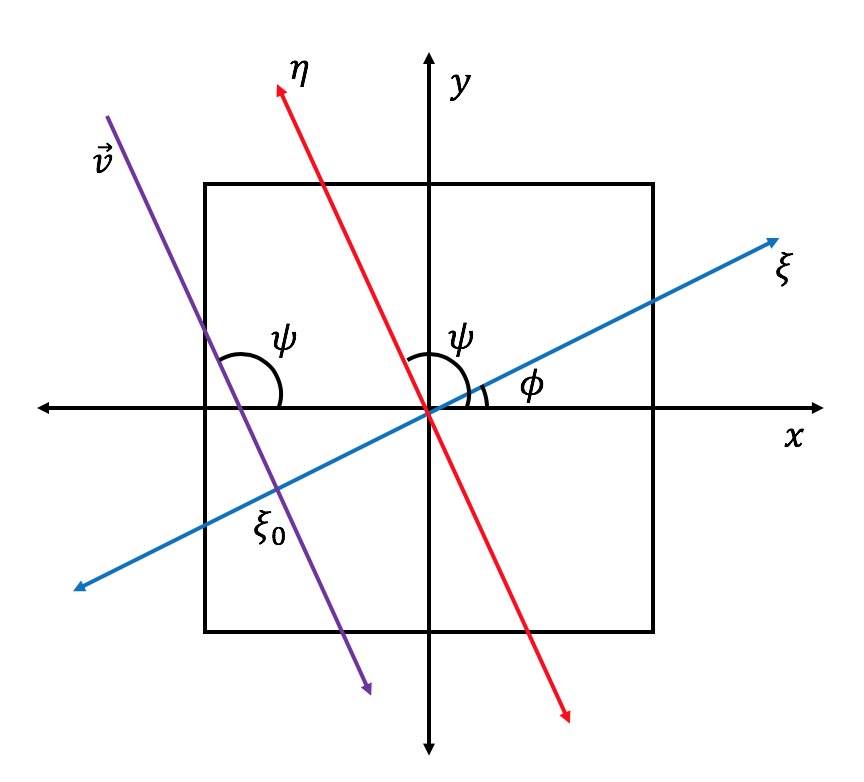}
        \captionsetup{width=1.1\linewidth, format=hang}
        \caption{\label{fig:ProjectionSpace}A diagram of projection space $\eta\xi$ from one incident muon with trajectory $\vec{v}$, angle $\psi$, traveling through the bottom slice of the pyramid shown in black, with the pyramid coordinate system $xy$ for reference.}
    \end{minipage}
    \hfill
    \begin{minipage}[t]{.52\textwidth}
        \includegraphics[width=1\textwidth]{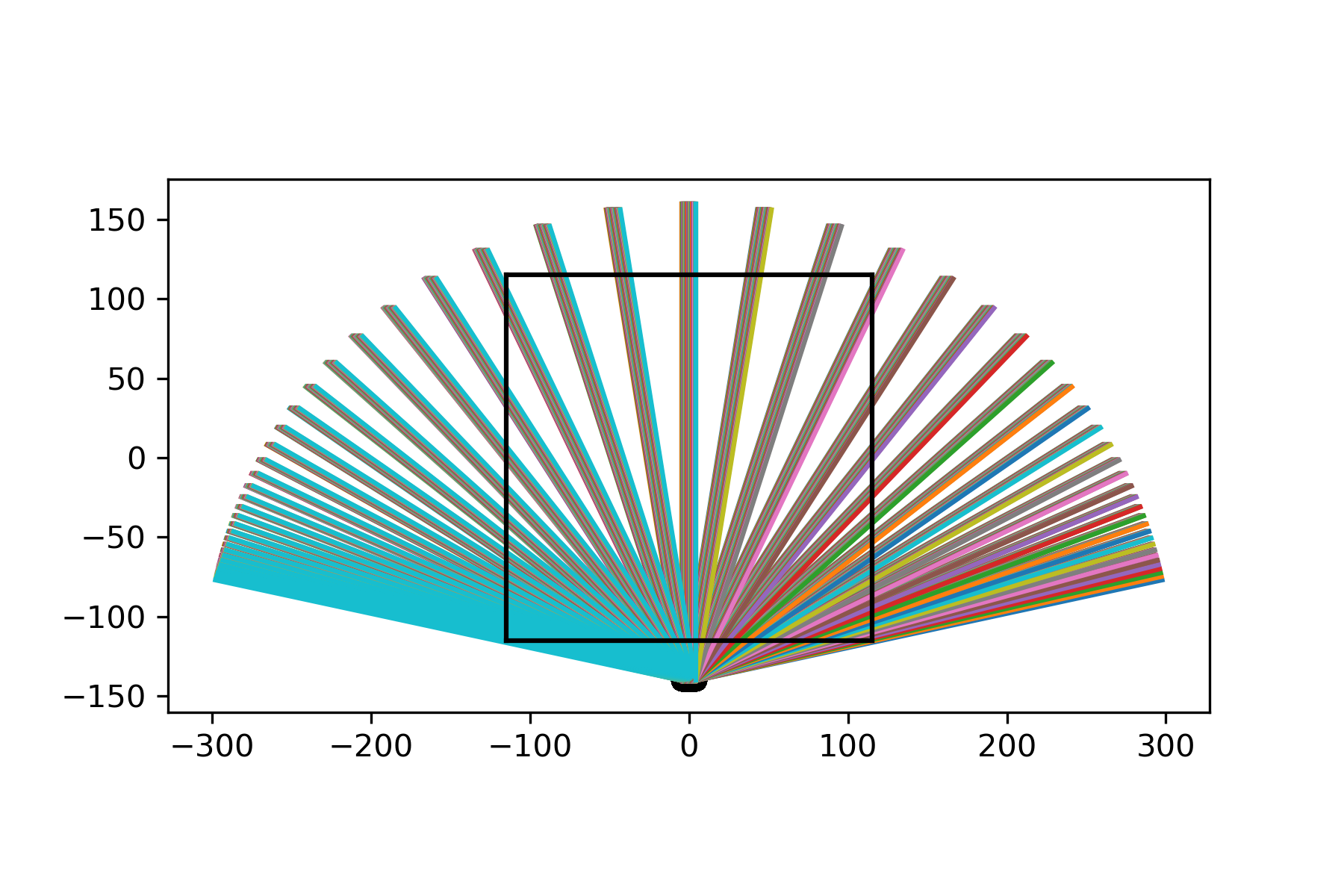}
        \captionsetup{width=1.1\linewidth, format=hang}
        \caption{\label{fig:FanDiagram}A diagram of the projection angles captures by every sixteenth pixel in a single row paired with every sixteenth pixel in the front panel of the same row. If every pixel is used the 'fan' is filled in.}
    \end{minipage}
    \centering
    \begin{minipage}[t]{.45\textwidth}
        \includegraphics[width=1\textwidth]{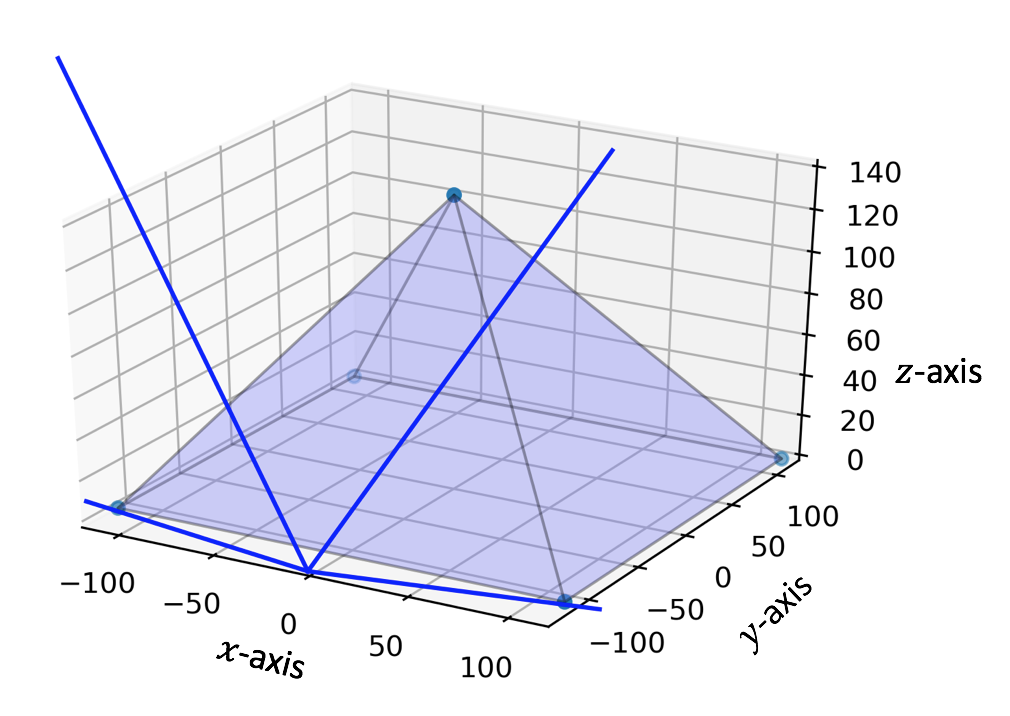}
        \captionsetup{width=1.1\linewidth, format=hang}
        \caption{\label{fig:DetectorRange}A plot of the range of trajectories with respect to the pyramid. If placed at the middle and $25m$ from the base of the pyramid, the entire pyramid falls within the trajectory range.}
    \end{minipage}
    \hfill
    \begin{minipage}[t]{.45\textwidth}
        \includegraphics[width=1\textwidth]{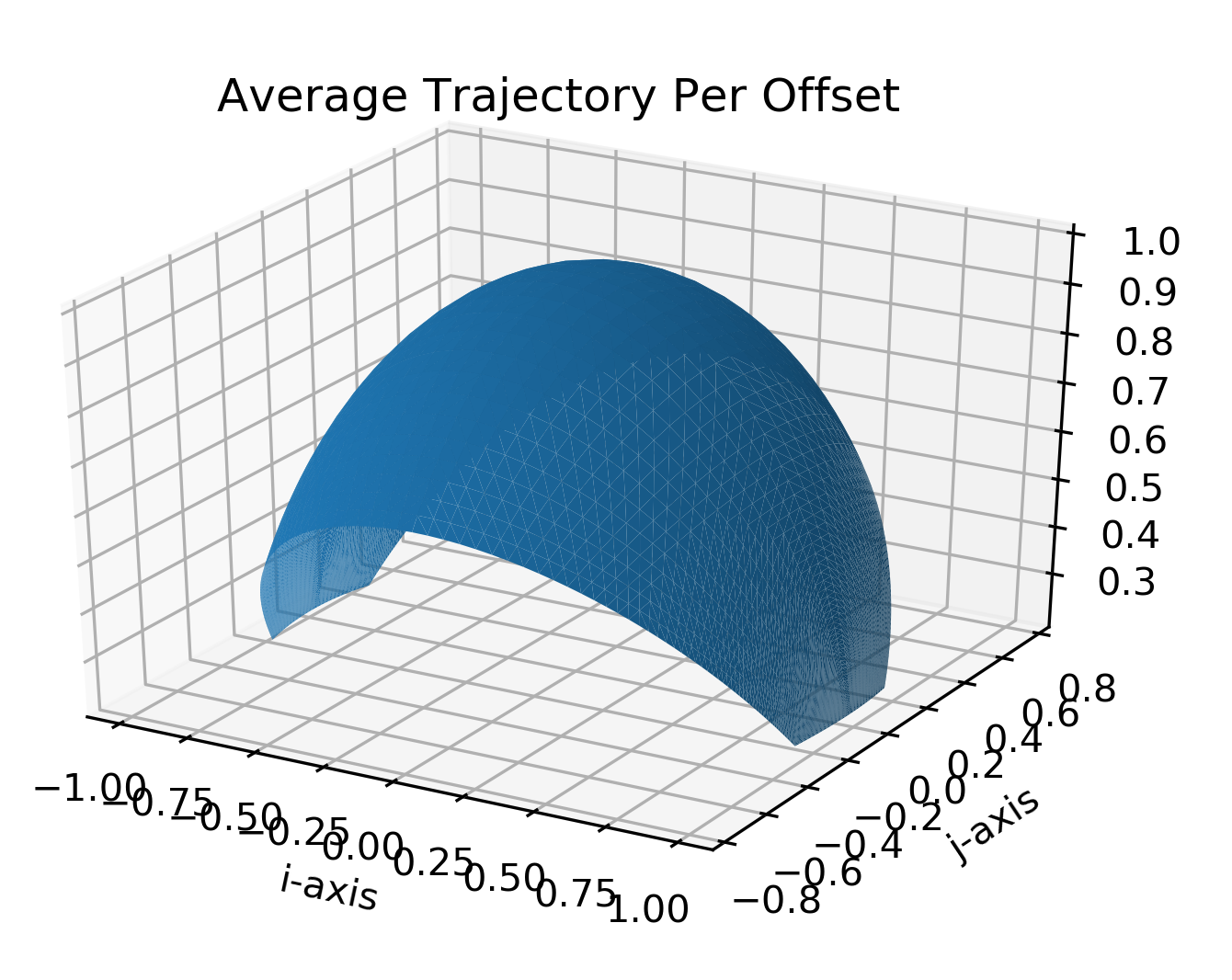}
        \captionsetup{width=1.1\linewidth, format=hang}
        \caption{\label{fig:DirectionofSight}A plot of the directions of sight as unit vectors using every eighth pixel.}
    \end{minipage}
\end{figure}

The direction of sight $r_{mn}$ is the direction of the vector passing through the center of two pixels. For pairs of pixels with the same horizontal and vertical displacement ($m = \Delta i, n = \Delta j$), the direction of sight is equal. The angular resolution for this particular direction of sight is given by the solid angle between these two pixels $\delta\Omega = A/r^2 = 4cm^2/r^2 $, as shown in Figure \ref{fig:SolidAngleDiagram}. Correcting for tilt, for each direction of sight $r_{mn}$ the angular resolution is given by the solid angle $\delta\Omega=\frac{A\cos\theta \cos\Psi}{r^2} $. This takes into account how the area covered by the solid angle formed by the connection of the far corners of offset pixels ($m = \Delta i \neq 0, n=\Delta j \neq 0$ is not equal to the area of the pixels, as shown in Figure \ref{fig:TiltedSolidAngleDiagram}.

With the solid angle as a function of the discrete directions of sight available from the detector, the number of pixel pairs that have that discrete direction of sight can be converted to the detection area available per solid angle. This is done with Equation 23 from Lesparre et. al.\cite{Lesparre}, $\tau(r_{mn})=S(r_{mn})\times \delta\Omega(r_{mn})$ representing Acceptance = Detection Area $\times$ Angular Resolution for each discrete direction of sight $r_{mn}$ with $m=\Delta i, n = \Delta j$. Doing this for every $r_{mn}$ returns the acceptance $\tau(r_{mn})$ of the detector.

\subsection{Sinogram Space}

Looking at one level of pixels, meaning all pixels of a row with a constant j $(\Delta j =0)$, the trajectories measurable through a 2d slice of an $xy$ layer of the pyramid at the height corresponding to $j$ can be calculated. The pyramid coordinate system is constructed with the origin at the center, $x$ parallel to the $i$-axis, $y$ parallel to the depth of the detector, and $z$ normal to the surface of the earth. From this single horizontal slice the angle of a trajectory is given by $\psi = \tan^-1(dy/dx)$ where $dy$ is equal to the $2m$ for all pixels due to the distance between the two detector plates, and $dx = \Delta i$. Converting from detector $ij$ space to pyramid $xyz$ space we can form projection space $\xi$ and $\eta$ with $\xi$ as the axis rotated by the azimuthal angle $\phi = \psi - \frac{\pi}{2}$\cite{Prince} around the $z$ axis. In projection space the angle of rotation $\phi$ as well as the intersection of the trajectory with the $\xi$ axis allows calculation of that trajectory's location in a sinogram. A plot of projection space  of one trajectory superimposed on the pyramid and pyramid coordinate system is shown in Figure \ref{fig:ProjectionSpace}.

With the left bottom corner of a panel being the origin of $ij$ space, and the angle of rotation of $\eta$ being $\psi=\phi + \pi/2$, $\xi = x\cos\psi +y\sin\psi$ and $\eta = -x\sin\psi + y\cos\psi$ with $\psi$ being the angle of the muon trajectory, i.e. the projection. The origin is shifted from the detector coordinate basis to the pyramid's coordinate basis with $x = (i - (960/2))/100$ and $y = -230.33/2 + 25$ both now in units of meters. With these conversions, for every possible muon trajectory representing a different angle and position of a projection, the sinogram space covered by the detector can be calculated. A plot of a subset of projection angles is shown in Figure \ref{fig:FanDiagram} as example of the range of projection angles for one row of the detector at one position.

Finally, as the amount of sinogram space covered by the detector in one position can be calculated, the detector can be moved and rotated in pyramid space and the amount of sinogram space covered by different positions, and multiple detectors, can be shown. As the pyramid is stationary the time of projection images captured can be ignored.  

\subsection{Resolution}

Lastly, the solid angle associated with each pixel pair was projected forward to its resolution at the center of the pyramid. The distance travelled and the resulting resolution is dependent on which vertical slice of the pyramid is imaged, with the center being the worst resolution assuming 360 degree rotation of the detector about the pyramid. For simplicity the lowest row of pixels was chosen, $j=0$, as it has the most relevant angle range, and that row was paired with every pixel on the front detector. This was done for one position of the detector with significant overlap of solid angles for different pairs of pixels with the same direction of sight.

\section{Results}

\subsection{Geometric Dimensions}
The geometric simulation of the maximum range of a stationary detector placed at $x=0$ and $25m$ from the base of the pyramid contained trajectories that encompassed the furthest edges of the pyramid. This is shown in Figure \ref{fig:DetectorRange}. The normalized direction of sight for every eighth pixel on the far panel with every eighth pixel on the front panel is shown in Figure \ref{fig:DirectionofSight} with the detector facing up. 

\subsection{Angular Resolution and Acceptance}
\begin{figure}[ht]
    \begin{minipage}[t]{.45\textwidth}
        \includegraphics[width=1\textwidth]{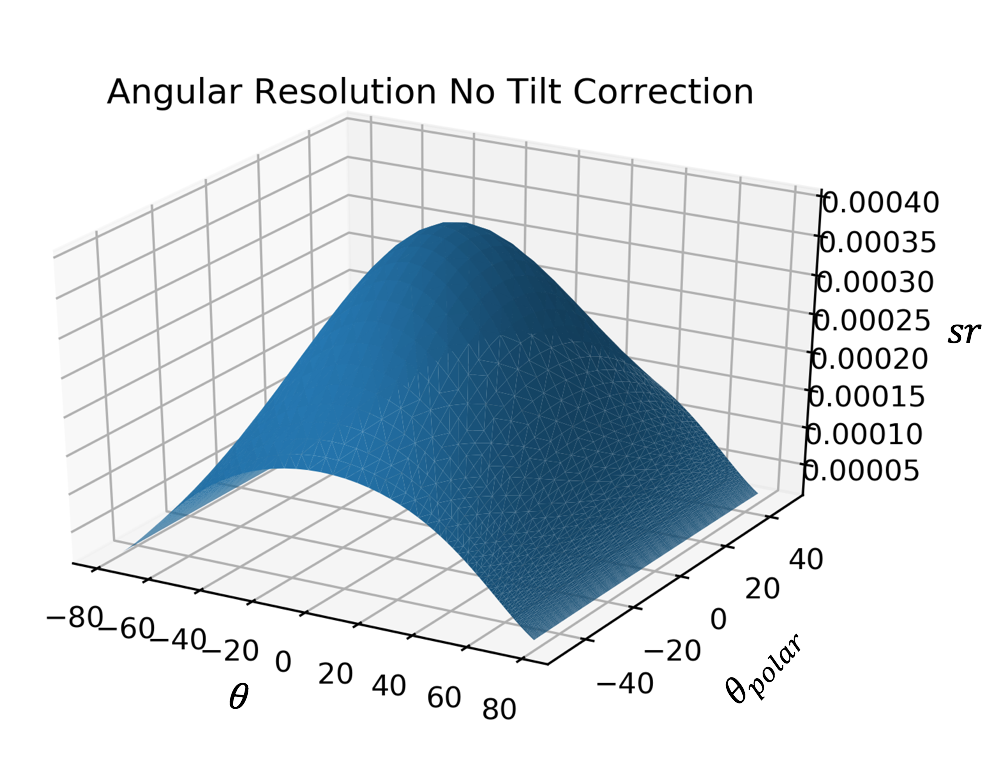}
        \captionsetup{width=1.1\linewidth, format=hang}
        \caption{\label{fig:AngularResolutionNoTilt}A plot of the angular resolution of the detector in $sr$ per angle of direction of sight. The effect of tilt on the solid angle is not taken into account.}
    \end{minipage}
    \hfill
    \begin{minipage}[t]{.45\textwidth}
        \includegraphics[width=1\textwidth]{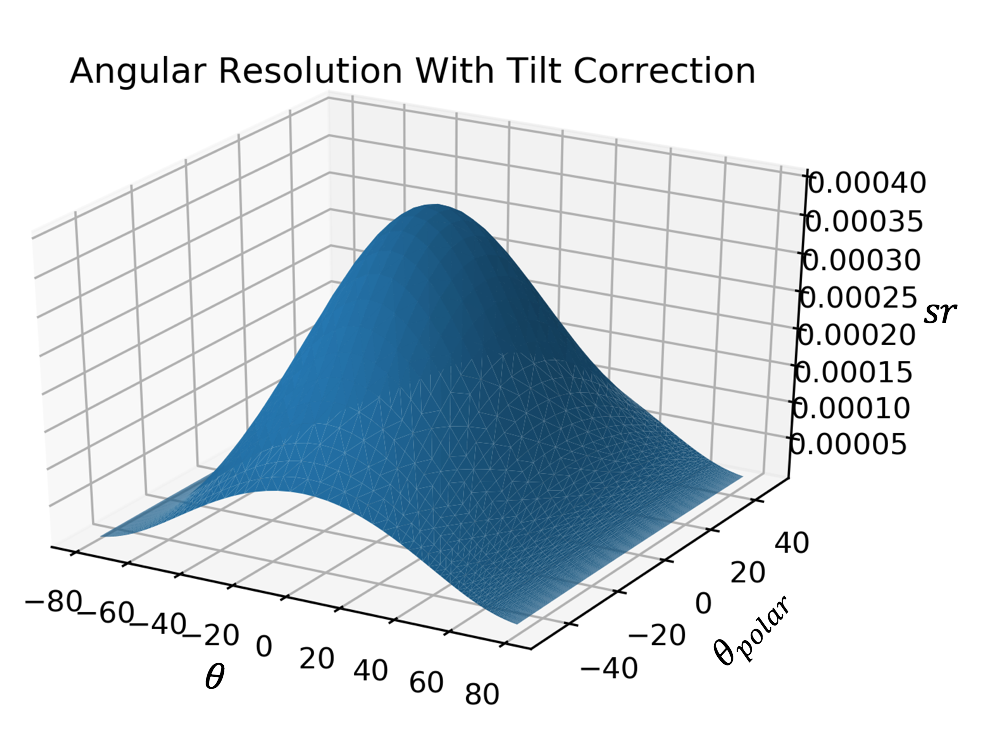}
        \captionsetup{width=1.1\linewidth, format=hang}
        \caption{\label{fig:AngularResolutionTilt}A plot of the angular resolution of the detector in $sr$ per angle of direction of sight. The effect of tilt on the solid angle is taken into account.}
    \end{minipage}
    \hfill
    \begin{minipage}[t]{.45\textwidth}
        \includegraphics[width=1\textwidth]{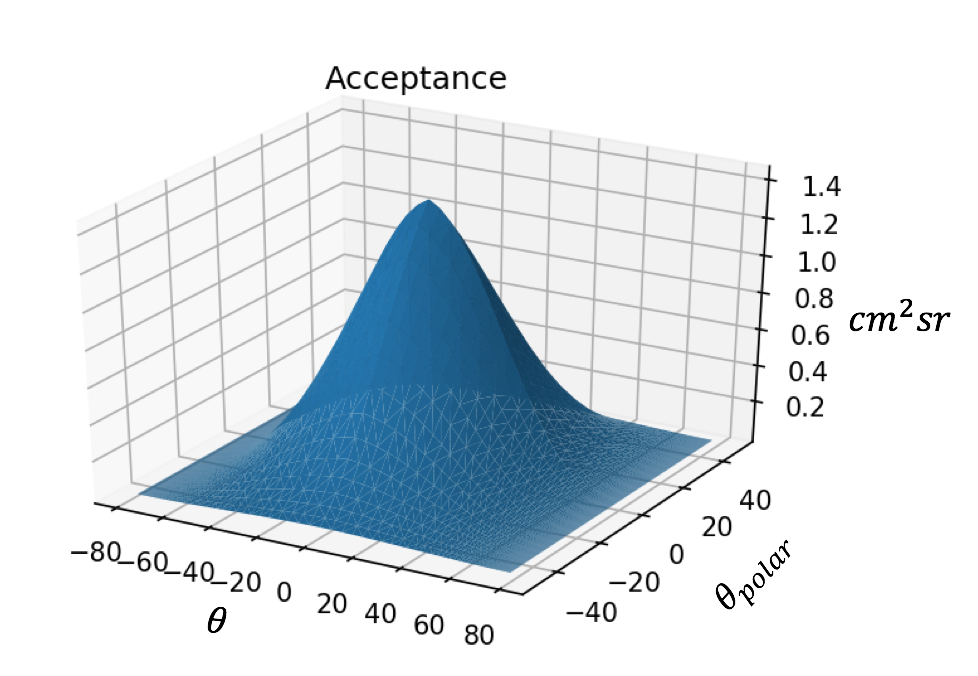}
        \captionsetup{width=1.1\linewidth, format=hang}
        \caption{\label{fig:Acceptance}A plot of the acceptance of the detector in $cm^2 ~sr$ per angle of direction of sight . The effect of tilt on the solid angle is taken into account.}
    \end{minipage}
    \hfill
    \begin{minipage}[t]{.45\textwidth}
        \includegraphics[width=1\textwidth]{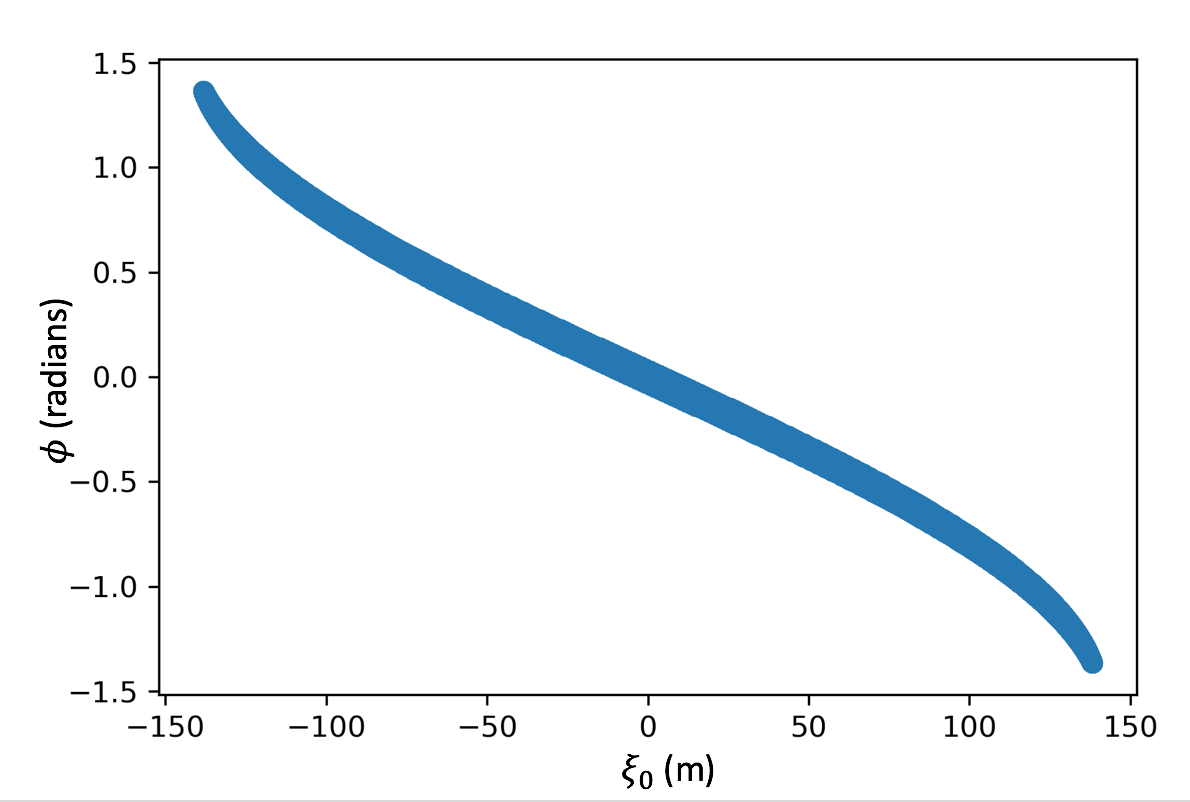}
        \captionsetup{width=1.1\linewidth, format=hang}
        \caption{\label{fig:SinogramPlot}A plot of the points in sinogram space covered by the detector located at the center of the $x$-axis and $25m$ from the base of the pyramid.}
    \end{minipage}
    \hfill
    \begin{minipage}[t]{.45\textwidth}
        \includegraphics[width=1\textwidth]{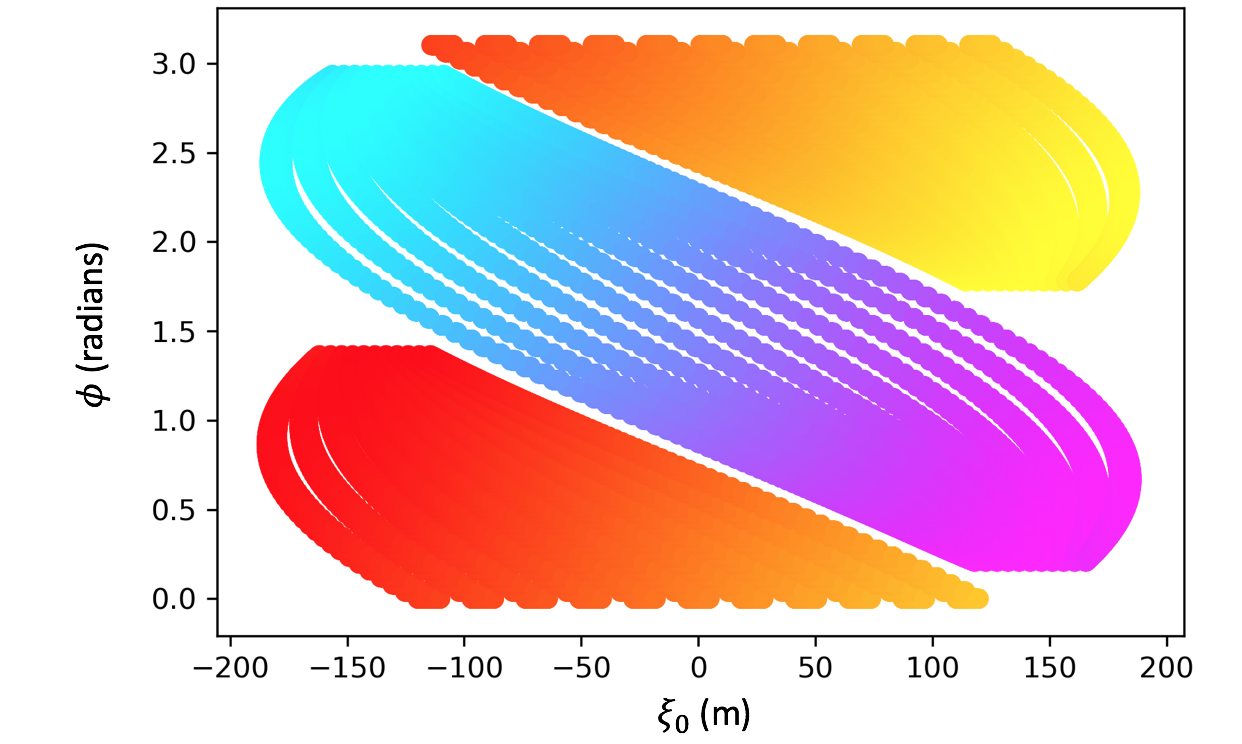}
        \captionsetup{width=1.1\linewidth, format=hang}
        \caption{\label{fig:SinogramCover}A plot of the points in sinogram space covered by a detector moved along the entire range of the pyramid along both the $x$ and $y$ axis eleven times ($~20m$ steps). In both cases the detector is pointed perpendicular to the axis it is travelling along, meaning it is facing the pyramid.}
    \end{minipage}
    \hfill
    \begin{minipage}[t]{.45\textwidth}
        \includegraphics[width=1\textwidth]{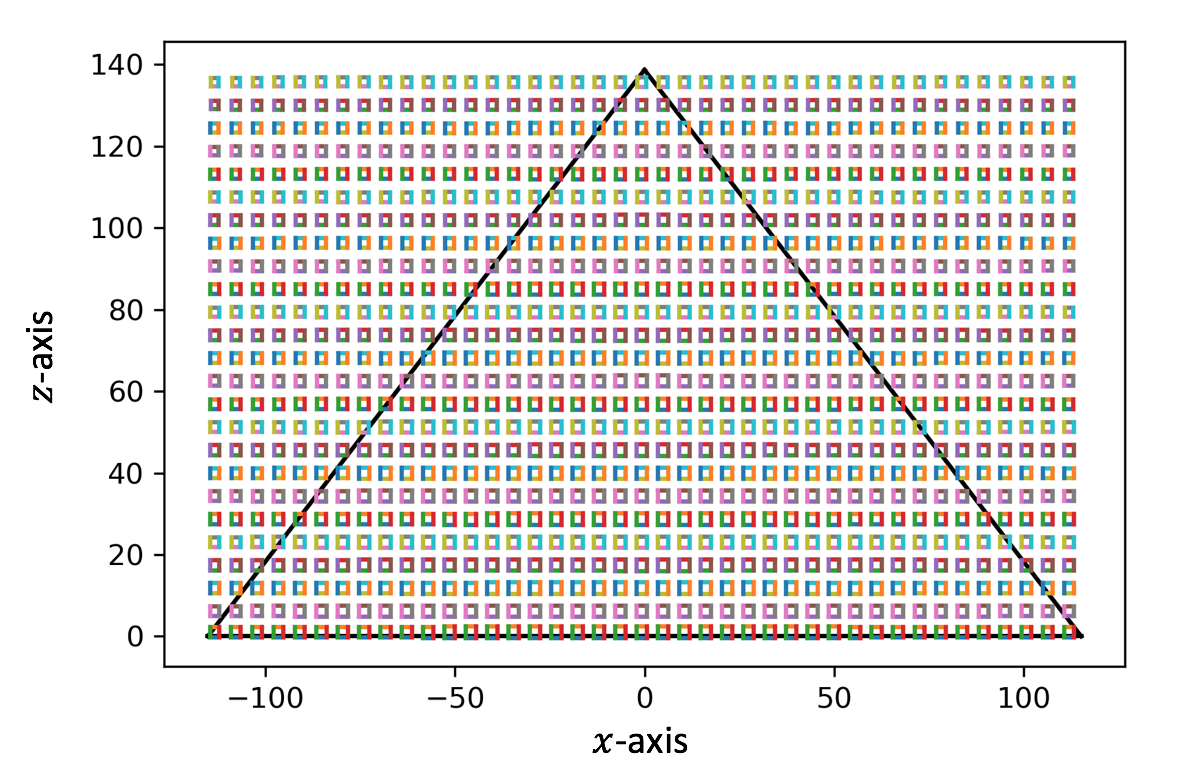}
        \captionsetup{width=1.1\linewidth, format=hang}
        \caption{\label{fig:Resolution}A plot of the resolution at the center slice of the pyramid for every eighth pixel for reference.}
    \end{minipage}
\end{figure}

The angular resolution, i.e. solid angle, associated with a certain direction of sight, is plotted as a function of $\theta$ and $\Psi$ without and with tilt correction as shown in Figures \ref{fig:AngularResolutionNoTilt} and \ref{fig:AngularResolutionTilt}. These values are as expected, as Lesparre et. al. finds a peak angular resolution of $.015 sr$ with $5 x 5 cm^2$ pixels and a $D =80cm$, because $\frac{(5\times 5)}{(80/2)^2)} = .015 sr$\cite{Lesparre}. Similarly, with the proposed detector with $2\times 2 cm^2$ pixels and a $D = 200 cm$, the peak angular resolution is $\frac{(2\times 2)}{(200/2)^2)} = .0004 sr$, which agrees with the simulation. Acceptance, i.e. detector area times angular resolution for each direction of sight, is shown in Figure \ref{fig:Acceptance}. The values calculated in Lespare et. al. are not matched by our detector. The peak acceptance calculated by hand for a $16 \times 16$ pixel grid with pixel size $25 cm^2$ with peak angular resolution of $.015 sr$ is equal to $16 \times 5^2 \times .015 = 100$. However the figure presented by Lesparre et. al. has a peak of $200 cm^2 sr$. The plot shown in Figure \ref{fig:Acceptance} follows the calculation done above, rather than the plot shown in the Lesparre paper.

\subsection{Sinogram Space}

The points of sinogram space covered by the detector at one position are shown in Figure \ref{fig:SinogramPlot}. This is reasonable as the width of the detector is 9.6m, while the width of the pyramid is 230m. Therefore at $\phi = 0$, completely forward direction of sight, the total range of $\xi_0$ covered is equal to the width of the detector. When a detector is moved in $20m$ increments along the $x$-axis of the pyramid at a constant $25m$ from the base, and another detector is moved along the $y$-axis of the pyramid, a majority of sinogram space is covered as shown in Figure \ref{fig:SinogramCover}.

\subsection{Resolution}
Lastly, the projections and the resulting angular resolution are shown as a grid at the center slice of the pyramid in Figure \ref{fig:Resolution}. There is a resolution of roughly $3\times 3 m^2$ which leads to an oversampling at every $2cm$ pixel. 

\section{Discussion}

This simulation showed the medical physics side of muon tomography and the possibility of high-resolution reconstruction using muon trajectories as tomographic line integrals. It also studied the angular resolution and acceptance of a current version of the detector, which can be changed based on resolution demands. The geometric simulation can also be used to test possible movements of detectors around the pyramid and the resulting sinogram space covered. The simulation shows that with detectors moving along two sides of the pyramid, a majority of sinogram space can be covered allowing possible reconstruction. The sampling resolution of sinogram space wanted for high-resolution reconstruction determines the step sizes between detector placement. The resolution at the center of the pyramid is roughly $3\times 3 m^2$, leading to oversampling.

The main limitation to the simulation is that it is purely geometric, does not take into account possible scattering, energy attenuation, the muon spectrum, timing, or  cosmic ray statistics. Future work involving statistical reconstruction involves taking into account coulomb scattering and scatter based on material properties as discussed by Schultz et. al \cite{Schultz} and Hanson et. al.\cite{Hanson}. Specifically the energy loss incurred by muons of initial energy $E_{0}>E_{min}$ travelling through the pyramid can be used to determine the opacity along the light of sight. That is, $E_0 - E_f = - \int_{0}^{L} l'\frac{dE}{dx}dl'=\int_{0}^{L}\rho(l')(\frac{dT}{\rho dx})_{E',l'} dl'$ as notated by Hanson et. al.\cite{Hanson}. This means that given the path length $L$, muons of a certain incident energy $E_0$ lose an amount of energy per density, and that knowing that path length one can know the opacity the muon travelled through to yield a final energy at the detector of $E_f$. The muon spectrum measured by the detector of open air could return a good estimate of $E_0$, and the change in the resulting spectrum per muon trajectory through the pyramid can be used to calculate the opacity of that trajectory. 

This simulation provides a way to calculate the positions needed to cover an adequate amount of sinogram space for high-resolution tomographic reconstruction, as well as a way to calculate the path length $L$ through the pyramid as a function of trajectory. In future work the energy attenuation as a function of opacity, comparison to open air, and scattering should be included. 

\section{Conclusion}

This simulation of geometric tracks of muon trajectories through a pyramid to a detector used in one-sided muon tomography showed the ability to fill sinogram space with a series of positioned detectors along the basis of the pyramid. It also showed the angular resolution and acceptance of the detector, and the resolution at the center of the pyramid. This suggests that using muon attenuation to calculate opacity of every trajectory one can reconstruct a high-resolution image of the pyramids using one sided tomography.

\section{Acknowledgments}
Supported, in part, by the U.S. National Institute of Biomedical Imaging and Bioengineering of the National Institutes of Health under grant number T32 EB002103.

\printbibliography


\end{document}